\documentclass[journal,transmag]{IEEEtran}

\usepackage{multirow}
\usepackage{caption}
\usepackage{subcaption}
\usepackage{tabularx}

\usepackage[hyphens]{url}
\usepackage[hidelinks]{hyperref}
\usepackage{amssymb}

\usepackage{booktabs,multirow}

\usepackage{xcolor}

\usepackage{amsmath}

\usepackage{subcaption}
\usepackage{booktabs} 

\usepackage{tabularx}

\usepackage{graphicx}

\usepackage{enumitem}
\usepackage{cite}
\usepackage{url}

\newcolumntype{L}[1]{>{\hsize=#1\hsize\raggedright\arraybackslash}X}

\begin{document}

\newcommand{\argmin}{\operatornamewithlimits{argmin}}
\newcommand{\argmax}{\operatornamewithlimits{argmax}}


\title
{AI-based Drone Assisted Human Rescue in Disaster Environments: Challenges and Opportunities}


\author{\IEEEauthorblockN{Narek Papyan\IEEEauthorrefmark{1},
Michel Kulhandjian\IEEEauthorrefmark{2}, Hovannes Kulhandjian\IEEEauthorrefmark{3}, and Levon Aslanyan\IEEEauthorrefmark{4}}
\IEEEauthorblockA{\IEEEauthorrefmark{1}Department of Telecommunication Engineering, Russian-Armenian University, Yerevan, Armenia}
\IEEEauthorblockA{\IEEEauthorrefmark{2}Department of Electrical and Computer Engineering, Rice University, Houston, TX, USA.}
\IEEEauthorblockA{\IEEEauthorrefmark{3}Department of Electrical and Computer Engineering, California State University, Fresno, Fresno, CA, USA.}
\IEEEauthorblockA{\IEEEauthorrefmark{4}Institute for Informatics and Automation Problems, National Academy of Sciences, Yerevan, Armenia}
\thanks{Manuscript received Oct. 11, 2023.}}




\IEEEtitleabstractindextext{
\begin{abstract}In this survey we are focusing on utilizing drone-based systems for the detection of individuals, particularly by identifying human screams and other distress signals. This study has significant relevance in post-disaster scenarios, including events such as earthquakes, hurricanes, military conflicts, wildfires, and more.

These drones are capable of hovering over disaster-stricken areas that may be challenging for rescue teams to access directly, enabling them to pinpoint potential locations where people might be trapped. Drones can cover larger areas in shorter timeframes compared to ground-based rescue efforts or even specially trained search dogs. Unmanned aerial vehicles (UAVs), commonly referred to as drones, are frequently deployed for search-and-rescue missions during disaster situations. Typically, drones capture aerial images to assess structural damage and identify the extent of the disaster. They also employ thermal imaging technology to detect body heat signatures, which can help locate individuals. In some cases, larger drones are used to deliver essential supplies to people stranded in isolated disaster-stricken areas.

In our discussions, we delve into the unique challenges associated with locating humans through aerial acoustics. The auditory system must distinguish between human cries and sounds that occur naturally, such as animal calls and wind. Additionally, it should be capable of recognizing distinct patterns related to signals like shouting, clapping, or other ways in which people attempt to signal rescue teams. To tackle this challenge, one solution involves harnessing artificial intelligence (AI) to analyze sound frequencies and identify common audio ``signatures''. Deep learning-based networks, such as convolutional neural networks (CNNs), can be trained using these signatures to filter out noise generated by drone motors and other environmental factors. Furthermore, employing signal processing techniques like the direction of arrival (DOA) based on microphone array signals can enhance the precision of tracking the source of human noises. 
\end{abstract}

}
\maketitle
\IEEEdisplaynontitleabstractindextext

\IEEEpeerreviewmaketitle


\section{Introduction}
\label{sec:intro}

Natural disasters represent dire situations that imperil the lives of numerous individuals. In these circumstances, telecommunication networks often suffer damage, impeding the effectiveness of search and rescue operations and necessitating significant time and resources for their restoration. In times of natural disasters and extreme weather conditions, search and rescue operations become paramount in mitigating risks to both human lives and the environment. Consequently, there is an urgent need to swiftly repair impaired telecommunication systems, enabling search and rescue teams to exchange critical information and facilitate real-time infrastructure recovery efforts.

The investigative process serves as a versatile system that enhances the efficiency of operations during extensive wildfires. Its advantage over existing methods lies in its ability to deliver quicker results. Continuous monitoring of the disaster can produce instantaneous outcomes, particularly by swiftly adjusting incident locations and coordinates, a task that surpasses human capabilities due to its high visibility and flexibility. Once the detection occurs, the coordinates are promptly decoded, data is analyzed, and processed information is transmitted to the appropriate institutions. Prior to the arrival of relevant rescue teams, taking the initiative to mitigate and minimize loss of life is imperative.

Continuous monitoring empowers us to maintain constant awareness and control over situations, enabling a responsive approach in every circumstance. This capability serves as a preventive measure against potential catastrophes. The implementation of this approach will involve the deployment of multiple drones, with a minimum requirement of at least two to ensure uninterrupted surveillance of the territory.




{\bf Related Work}: Several works have been undertaken in the field of search and rescue operations.
Alsamhi \textit{et al.} \cite{alsamhi2022uav}, assess the network performance of UAV-supported intelligent edge computing to accelerate Search And Rescue (SAR) operations. This technology offers rapid deployment capabilities and can aid in disaster-related rescues.

Alawad \textit{et al.} \cite{alawad2023unmanned}, propose a crisis and disaster management system that leverages a swarm optimization algorithm (SOA) to enhance disaster and crisis management efforts. Within this system, the UAV search and rescue team operates under the delay tolerant network (DTN) strategy, enabling efficient exploration capabilities.


Erdelj and Natalizio \cite{erdelj2016uav}, identify the primary applications of UAV networks in disaster management and examine key research challenges within the domain of UAV-assisted disaster management.

In our previous work, \cite{kulhandjian2023ai}, we have created an integrated system featuring two distinct sensing mechanisms, enabling real-time detection and precise localization of humans and animals even in thick smoke, enhancing the situational awareness of firefighters on-site. An extended version of this work has been submitted for a patent \cite{kulhandjian2022system}. 

In this article, we illustrate our strategy for combatting large-scale wildfires through the utilization of unmanned aerial vehicles (UAVs) and neural network algorithms, with a primary focus on significantly reducing human casualties \cite{Albanese2021}. UAV technology plays a pivotal role in mitigating the impact of emergency situations and expediting rescue and recovery efforts \cite{Namuduri2017, Tomic2012, Grover2017}. Our foremost objective is the swift detection of emerging disasters and prompt reporting to all relevant law enforcement agencies. A swift response to such incidents presents a valuable opportunity to contain the wildfire's spread, with every moment gained being crucial in limiting the disaster's extent. 


Moreover, we conduct a comprehensive exploration of speech recognition algorithms, perform comparative assessments, and uncover their potential applications in diverse environmental conditions. Our UAVs are outfitted to receive audio signals and discern the source's orientation via an advanced microphone array.

Once the information is received, we initiate signal processing using specialized neural network algorithms designed for speech recognition. After extensive testing with UAVs, we have verified the efficiency of microphone array technology, which depends on sound and voice recognition algorithms. This technology accurately pinpoints the direction and location of sound sources.

In the monitoring process, the use of cameras and advanced image processing algorithms, leveraging deep neural networks, enables us to pinpoint the origins of the disaster, evaluate its scale, and approximate the initial coordinates of the incident location.

After the initial detection of the disaster, to obtain more precise location coordinates, we will utilize techniques for detecting, analyzing, and processing audio signals. Employing a microphone array, we will capture specific information related to the disaster's location. Subsequently, this data will be processed using deep learning algorithms, including speech recognition techniques.

In our research, we harnessed two distinct neural network algorithms—one crafted for image processing and the other for sound recognition. Each of these algorithms serves a unique purpose, and their combined utilization allows us to complement and address each other's limitations effectively. 

The rest of the paper is organized as follows. In Section \ref{UAVAssistedDisaster}, we present UAV-assisted disaster management with applications, followed by the situation awareness and logistics and evacuation support in Section \ref{SituationAwareness}, and with the SAR missions in \ref{SARMissions}.  
In Section \ref{MedicalMaterialDelivery}, 
   we discuss the medical material delivery, telemedicine platform application.
 In Section \ref{State-of-The-ArtTechnologies},  we present the state-of-the-art technologies using UAVs to localize people in emergency situations.
In Section \ref{AdvantagesAndDisadvantages},  we present the advantages and disadvantages of the hardware, followed by design challenges for UAV-assisted emergency networks in Section \ref{DesignChallenges} and the design architecture in Section \ref{DesignArchitecture}.
 Finally, we present future directions in Section \ref{FutureDirections} and draw the main conclusions in Section \ref{conclusion}.

\section{UAV Assisted Disaster Management Applications}
\label{UAVAssistedDisaster}
\subsection{Monitoring, Forecasting, and Early Warnings}

Environmental monitoring and the proactive identification of potential emergency situations constitute the primary line of defense in protecting the population from both natural and human-induced hazards. Our capacity to rapidly and effectively locate sources of risk empowers us to anticipate potential disruptions within specific regions and pre-emptively recognize imminent threats. This forward-looking strategy equips us to put in place all essential safeguards, preventing these hazards from evolving into full-fledged emergencies. The accuracy of our predictions related to weather conditions, natural calamities, or human-generated mishaps directly influences their effectiveness in preserving human lives and welfare.
 
To address these challenges, a range of monitoring, data analysis, and processing methods are employed. The contemporary system for emergency monitoring and forecasting is designed to observe, supervise, and predict hazardous weather events and natural occurrences, as well as technological developments and their evolving dynamics. Emergency forecasting serves the purpose of assessing the magnitude of potential disasters and orchestrating efficient preventive measures in response to them. 
 
Utilizing UAVs, we can effectively search for individuals who are lost in forests, inspect remote and challenging-to-access areas, and carry out other essential operations. Nevertheless, drones do have a notable limitation: their flight duration. To overcome this constraint and cover extensive areas while gathering the additional data required for modeling and predicting various emergency situations, we will address this issue by deploying multiple UAVs instead of relying on a single one.

\begin{enumerate}[label=(\arabic*)]
    \item Monitoring: Our primary objective involves the use of a surveillance camera mounted on a drone. By leveraging the capabilities of the Arducam IMX477 camera, we will maintain constant awareness of the ongoing situation and have the capacity to document even the slightest alterations. Furthermore, the thermal camera ``Seek Thermal'' ensures that visibility remains unimpeded during nighttime events.
    
    \item Forecasting: Following the reception of data collected by the cameras, this phase will commence. The information will be transmitted to the Jetson Nano microcomputer, which will undertake image processing using advanced deep-learning techniques. The Jetson Nano microcomputer is renowned for its superior performance, especially when working with artificial intelligence algorithms, outperforming even the Raspberry Pi. Upon completion of the algorithm, we will obtain a forecast indicating the presence of a disaster and its likelihood.
    \item Early warning: This constitutes a crucial juncture, as it is the very reason for conducting monitoring. Without this step, the purpose of terrain exploration and forecasting would be rendered meaningless. Upon obtaining the probability forecast, we will promptly juxtapose it with our predefined threshold value. If the result surpasses this threshold, we will expeditiously trigger an event and transmit the pertinent information to the relevant authorities, ensuring no time is wasted in taking necessary actions.
    The event will encompass the following information: 
    \begin{itemize}
    \item The event will include the time at which the disaster was detected. (day/month/year: hour/minute/second)
    \item The event will provide precise coordinates, along with details regarding the scope and scale of the fire.
    \item Additionally, the event will convey information about the direction in which the fire is spreading.
    \item Furthermore, the event will furnish data concerning the speed at which the fire is advancing.
    \end{itemize}

\end{enumerate}
The elements mentioned encompass all the essential details required by the relevant authorities. Armed with information regarding the scale and direction of the incident, responders can act swiftly and efficiently to prevent and extinguish the fire.

The precision of monitoring and forecasting emergency situations significantly impacts the effectiveness of mitigating the risks associated with their occurrence, propagation, and suppression of potential catastrophes that have not yet fully developed.

\subsection{Disaster Information Fusion and Sharing}
In responding to emergency situations, the integration of data from multiple sources is of paramount importance, as it enables the delivery of the most up-to-date, precise, and timely information across various scales to support disaster risk reduction services.

Information integration involves amalgamating data from diverse sources, each with varying conceptual, contextual, and typographic representations. This process is employed for tasks such as data mining and consolidating information from unstructured or semi-structured resources. While information integration primarily pertains to textual knowledge representation, it can also extend to multimedia content. Noteworthy technologies utilized for information integration encompass deduplication and string metrics, which facilitate the identification of similar text across different data sources through fuzzy matching. Additionally, some methods rely on causal outcome estimates based on a source model to enhance the integration process.

Information fusion and sensor fusion, often referred to as multi-sensor fusion, are interconnected concepts related to information integration.

Information fusion encompasses the process of amalgamating data from multiple sources to generate a new dataset that accurately represents a true value for a particular data item, especially when different data sources offer conflicting information. A variety of algorithms have been developed to address this challenge, spanning from straightforward techniques like majority voting to more intricate approaches capable of assessing the reliability of various data sources. These methods play a vital role in enhancing the accuracy and reliability of integrated information for decision-making and problem-solving in various fields, including sensor networks and data analysis.

Sensor fusion is the procedure of merging data obtained from sensors or diverse sources to produce information that has reduced uncertainty compared to using these sources individually. This approach aims to achieve a result that is not only ``more accurate'' but also ``more comprehensive'' when compared to relying on a single sensor.

For instance, it is conceivable to enhance the precision of determining the location of an object within an environment by integrating data from multiple sources, such as video cameras, GPS systems, and other sensor inputs. In our project, we will promptly integrate data from GPS, cameras, and microphone arrays to improve the overall accuracy and completeness of our information gathering and analysis. 

\begin{figure*}[h]
\centering
\includegraphics[width=12.7cm]{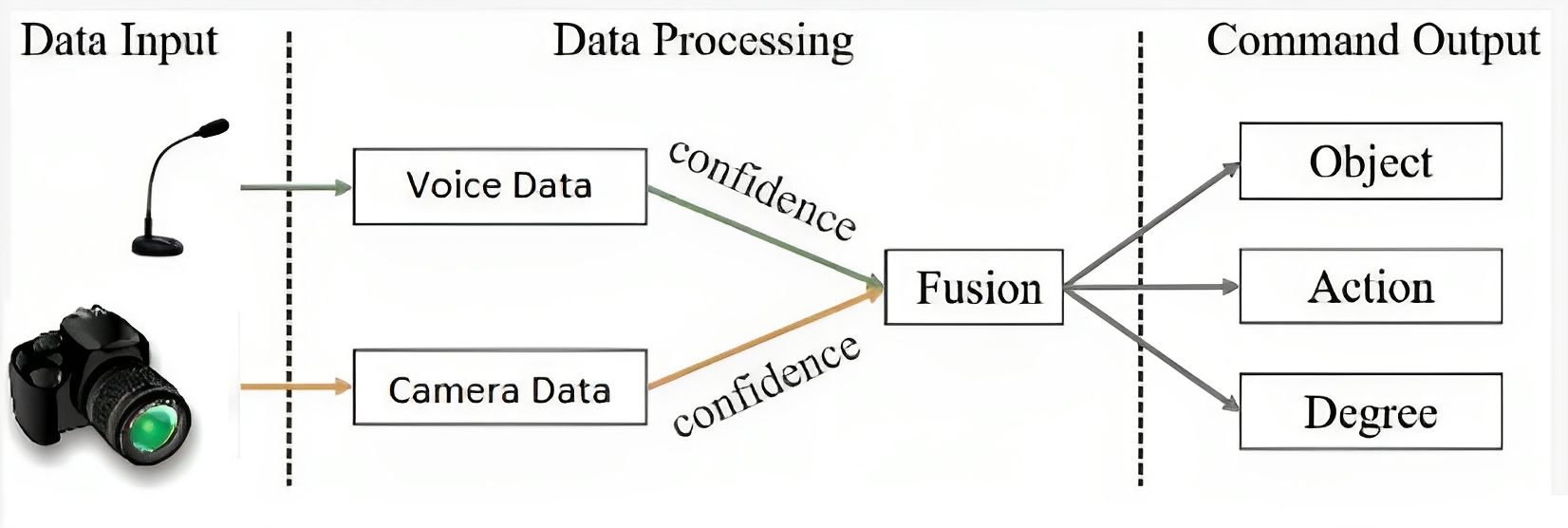}
\caption{Data Processing.}
\label{DataProcessing}
\end{figure*} 

This method provides a significant advantage when utilizing data collected from separate sources, compared to relying solely on their individual results. Both camera and sound data processing yield outcomes in the form of probabilities, enabling us to determine a positive result when our probability exceeds a predefined threshold. For instance, if our minimum threshold for both the camera and microphone is set at 0.8, any values below this threshold will be disregarded. When using the data sources independently, any value below 0.8 would result in a negative determination.

As an example, if the camera data yields a probability of 0.7 and the microphone data yields 0.6, under standard circumstances, we would conclude that there is no event. However, through the information integration method, we depart from the conventional approach and combine the results, resulting in 0.7 and 0.6 becoming components of a single assessment. In this way, we arrive at the truth of the event, a conclusion that would not be attainable without the application of information integration principles. This approach allows us to leverage the strengths of multiple data sources and increase the reliability of our results.

\section{Situation Awareness and Logistics and Evacuation Support}
\label{SituationAwareness}
Situational awareness (SA) entails perceiving elements and events in the environment concerning time and space, comprehending their significance, and anticipating their future developments. At its core, SA involves understanding what is transpiring in the environment and how it impacts both the present and the future. While this may appear straightforward in stable and controllable settings, it can pose substantial challenges in swiftly evolving and intricate circumstances. Consequently, situational awareness is particularly pertinent in scenarios characterized by a high degree of variability, uncertainty, complexity, and ambiguity. One of the principal objectives of situational awareness is the identification and prevention of errors, making it an indispensable concept in various domains where precision and adaptability are crucial.

The formal definition of SA can be divided into three key components: the perception of elements within the environment, the comprehension of the situation, and the ability to forecast future developments. Research in the field of SA has primarily focused on three aspects: States of SA, Systems of SA, and Processes of SA.

\begin{enumerate}
\item \textbf{States of SA}: This aspect pertains to the actual level of awareness an individual possesses regarding a given situation.

\item \textbf{Systems of SA}: This aspect concerns the distribution of SA among objects within the environment and the exchange of SA between various components of a system.

\item \textbf{Processes of SA}: This aspect involves the continuous updating of SA states and the factors that drive instantaneous changes in SA.
\end{enumerate}
Numerous models have been developed to describe and understand situational awareness, with one of the most widely recognized and practical models being the three-level Endsley model. This model provides a structured framework for comprehending and assessing situational awareness in complex environments.

\subsection{Perception of SA}
The first level of situational awareness is closely linked to the perception of pertinent information. This has two significant implications. Firstly, individuals must have access to relevant information, which they must recognize and grasp within seconds of accessing it. Consequently, one of the fundamental prerequisites for achieving the first level of SA is effective communication and proper visualization. For instance, if a project manager is not informed about a potential issue that could lead to project delays, they cannot initiate corrective actions to address it. Consequently, they lack the first level of SA, which hinders their ability to make informed decisions and steer the project in the right direction.

\subsection{Understanding the SA}
The second level of SA involves the essential task of accurately comprehending pertinent information. Depending on the specific situation, this necessitates having the appropriate knowledge to effectively interpret the received information.

Mental models play a pivotal role in achieving the second level of SA. This is because individuals construct new mental models or modify existing ones based on how they interpret information. When crucial information is absent or the data is incorrect, it implies that the mental model is flawed. Consequently, individuals encounter difficulties at the second level of SA, as their understanding of the situation is compromised. 

\subsection{Forecasting}
The third and final level of SA pertains to the ability to predict future states based on perceived and relevant information. This becomes especially crucial when dynamic processes are anticipated in the future, often relying on assumptions.

In complex systems, marked by a high degree of interdependence, it becomes challenging to predict how changes in one variable might influence the overall state of the system. This underscores the significance of accurate forecasting at the third level of SA.

Moreover, incorrect or outdated mental models can also lead to shortcomings at the third level of SA, as they can result in flawed assumptions about future developments and hinder the ability to make well-informed decisions in rapidly evolving situations.

\section{SAR Missions – WAVs Can Search Dor and Rescue People Lost, Injured or Trapped by Debris}
\label{SARMissions}
Search and Rescue (SAR) is a voluntary non-profit organization dedicated to locating and rescuing individuals who are lost, missing, or injured, particularly in outdoor and street-related emergencies. SAR is an extensive emergency service, involving highly trained military specialists, local law enforcement agencies, and dedicated civilian volunteers. The primary objective of SAR is to identify, provide initial medical care, and safely evacuate individuals in distress.

Efficiently saving lives during and after an incident or a natural disaster is a time-consuming and perilous task for rescue teams. The longer the rescue operation takes, the greater the risk to those in need. In such critical scenarios, drones offer a ray of hope. They significantly reduce the time required to search for individuals or objects, decrease operational costs, and, most importantly, minimize the risks faced by search and rescue personnel.

\section{Medical Material Delivery/Telemedicine Platform}
\label{MedicalMaterialDelivery}
The future use of drones in healthcare presents a range of opportunities to enhance safety and healthcare delivery. Here's how the industry can best utilize this technology:

\begin{itemize}
\item \textbf{Disaster Response and Relief}: Drones can play a critical role in delivering food aid and medical supplies to disaster-stricken areas. Rapid delivery of essential items directly to disaster zones can help prevent outbreaks of life-threatening infectious diseases. This includes delivering communication equipment, mobile medical units, and portable shelter, particularly in situations where damage to critical infrastructure disrupts ground or conventional air transportation.

\item \textbf{Remote Medical Care}: Drones facilitate more effective medical care for patients in remote or mobile settings. They can deliver medicines and supplies to patients receiving home care instead of being in a hospital. For example, when a healthcare worker visits a homebound patient, blood samples can be collected and sent by drone to a laboratory for analysis. Medications, antibiotics, and prescribed treatments can be delivered to patients' homes. This technology can extend the duration of home care for individuals in nursing homes, enhancing the independence of the aging population. Drones can also monitor patients with conditions like dementia or deliver food to those who cannot cook independently.
\end{itemize}
Incorporating drones into healthcare operations not only expedites critical responses in emergencies but also expands access to medical services for patients in remote locations or receiving home care. It has the potential to revolutionize healthcare delivery by making it more efficient, convenient, and responsive to the needs of diverse patient populations.

Radar technology is used to determine physiological parameters such as breathing and heart rate by emitting periodic, known narrowband pulses that interact with a person. The time delay and frequency modulation in the received signal are directly related to the target's unknown range and speed, respectively. By analyzing these parameters in the received signal, radar systems can deduce critical health information. This includes detecting the micro-Doppler effect caused by tiny movements of body parts like the lungs and heart; for example, breathing causes millimeter-level displacement, while heartbeat results in sub-millimeter displacement.

In healthcare, drone systems have the potential to address a wide range of challenges faced by healthcare workers, from emergency responders in underserved regions to busy hospital staff. Drones can rapidly transport vital medical supplies like blood, vaccines, contraceptives, and antivenom to remote areas, reaching patients in need of immediate care within minutes, which can be a life-or-death situation.

They can also streamline logistics within hospital premises, transferring medications between different hospital buildings and providing tools to assist elderly patients in aging in place. Drones offer numerous promising opportunities for the healthcare sector, not only in terms of saving lives but also in reducing costs.

One significant advantage of drones in healthcare is their ability to cover longer distances at higher speeds when transporting blood products and laboratory samples. Currently, ground vehicles are often used for such transport, which can result in accidents and delays. Drones can mitigate these issues and enhance the efficiency of healthcare logistics.

However, like any industry adopting drone technology, healthcare faces various challenges, including payload capacity, battery life, and regulatory compliance. Addressing these issues will be crucial for realizing the full potential of drones in healthcare.

\section{State-of-The-Art Technologies Using UAVs to Localize People in Emergency Situations}
\label{State-of-The-ArtTechnologies}
\subsection{Sensors
Accelerometer, Magnetometer, Gyroscope, Barometer
}

A sensor is a device designed to detect and respond to various types of input from the physical environment. These inputs can encompass a wide range of phenomena, including light, heat, motion, humidity, pressure, and numerous other environmental factors. Sensors serve as a crucial link between the physical world and the digital realm by essentially acting as the ``eyes and ears'' for computing systems. They collect data from their surroundings, which is then analyzed and acted upon by the computing infrastructure.

Sensors can be categorized in several ways, one of which is the division between active and passive sensors:

\begin{itemize}
\item \textbf{Active Sensor}: An active sensor is a type of sensor that necessitates an external power supply to respond to environmental inputs and generate outputs. For instance, sensors used in meteorological satellites require a power source to provide meteorological data about the Earth's atmosphere.

\item \textbf{Passive Sensor}: In contrast, a passive sensor doesn't rely on an external power source to detect environmental influences. It harnesses the energy present in the environment itself, such as light or heat energy, to function. A classic example is the mercury glass thermometer, where temperature changes cause the level of mercury in a glass tube to rise or fall, providing an easy-to-read gauge for temperature measurement.
\end{itemize}
Certain sensor types, like seismic and infrared sensors, are available in both active and passive forms, offering versatility in various applications.

Another classification criterion for sensors is whether they are analog or digital, based on the type of output they produce:
\begin{itemize}
\item \textbf{Analog Sensor}: Analog sensors convert environmental inputs into analog outputs, which are continuous and variable. These sensors provide data in a continuous manner, representing a range of values.

\item \textbf{Digital Sensor}: Digital sensors, on the other hand, convert environmental inputs into discrete digital signals, which are transmitted in binary format (comprising 1s and 0s). Digital sensors provide data in a binary, on/off manner, allowing for more precise and straightforward data processing and transmission.
\end{itemize}
The choice between analog and digital sensors depends on the specific application and the requirements for data accuracy, resolution, and processing.

\subsection{Video Camera/RGB Depth Camera}
The term ``RGB'' refers to a color model that combines the primary colors of light, namely red, green, and blue, to produce a wide range of colors that humans perceive. An RGB camera is a type of camera used to capture color images by recording light in these red, green, and blue wavelengths (RGB). This camera operates within the visible light spectrum, typically ranging from 400 to 700 nanometers (nm).

In contrast, an RGBD camera is a specialized type of depth camera that provides both depth (D) and color (RGB) data in real time. The depth information is obtained from a depth map or image generated by a 3D depth sensor, such as a stereo or time-of-flight sensor. RGBD cameras merge RGB color data and depth information at a pixel level, allowing them to convey both types of data within the same frame.

RGBD cameras are highly preferred in certain embedded vision systems for several reasons:

\begin{enumerate}
\item \textbf{Enhanced Object Identification}: Combining RGB color data with depth information enables more efficient and accurate object recognition and pattern detection. This is particularly valuable in applications that require identifying and characterizing objects within a scene, as well as measuring their distance from the camera.

\item \textbf{Applications}: RGBD cameras find applications in various fields, including anti-spoofing systems based on face recognition and people-counting devices, where the ability to capture both color and depth information simultaneously is critical for accurate results.
\end{enumerate}

\begin{figure}[b]
\centering
\includegraphics[width=4.5cm]{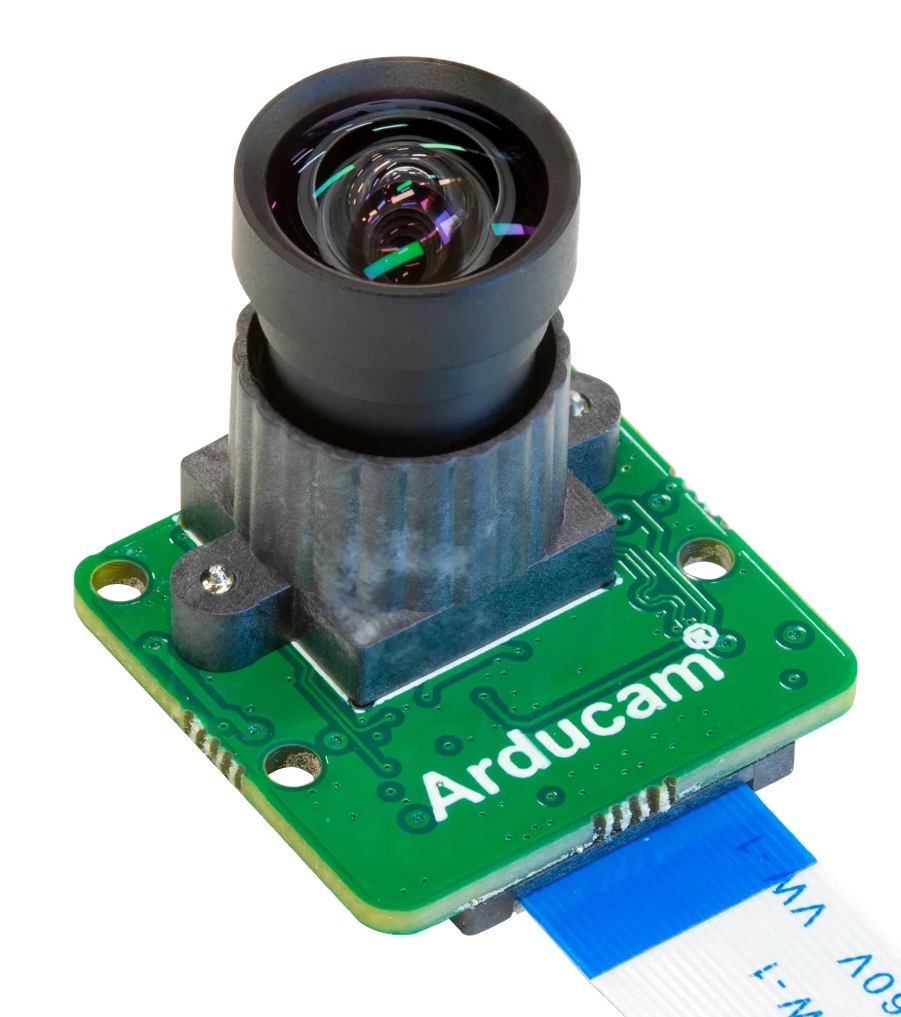}
\caption{Arducom camera.}
\label{Arducom}
\end{figure}

\begin{figure}[b]
\centering
\includegraphics[width=6.0cm]{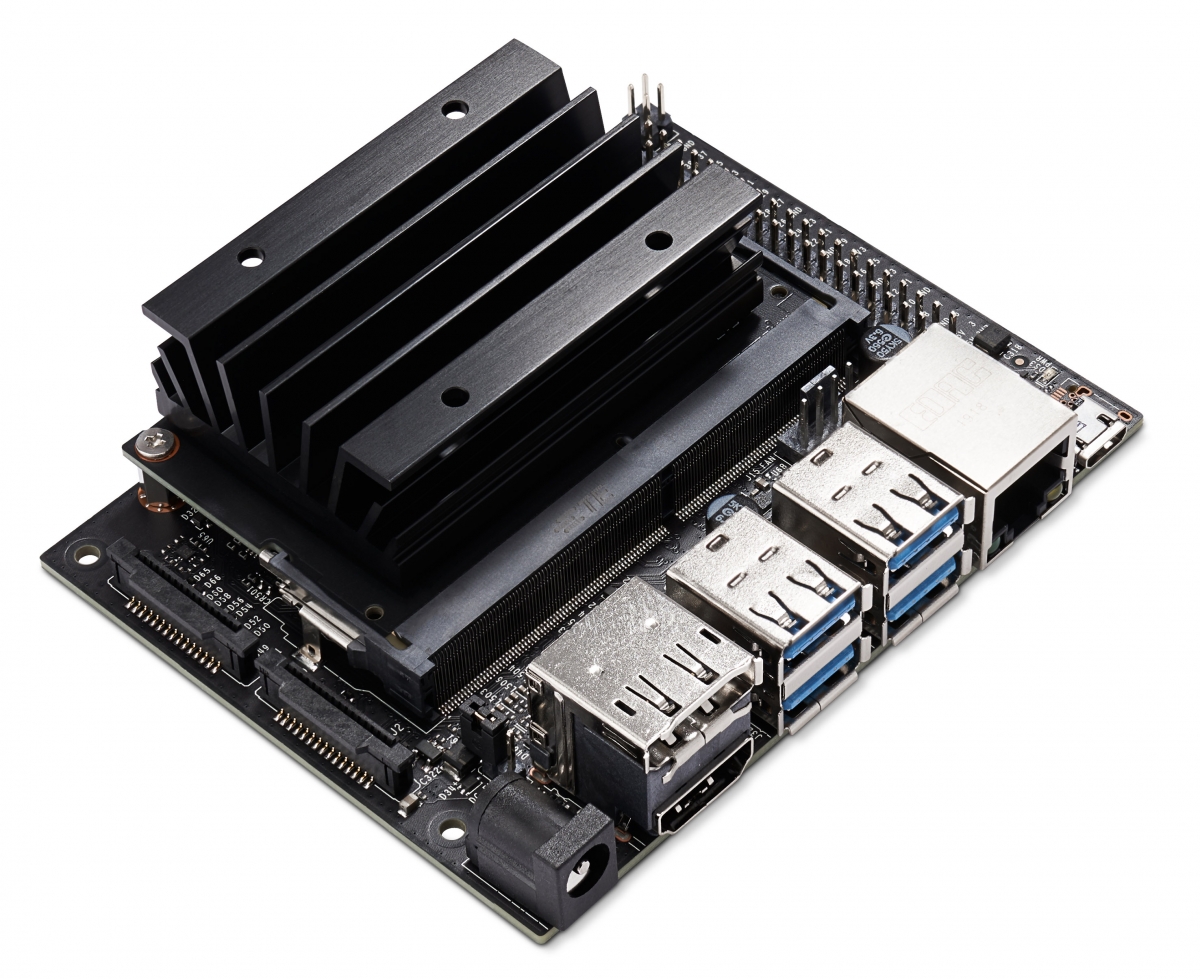}
\caption{Nvidia Jetson Nano.}
\label{Jetson}
\end{figure}

The Arducam IMX477 synchronized stereo camera kit shown in Fig. \ref{Arducom} is designed for use with the Nvidia Jetson Nano which is portrayed in Fig. \ref{Jetson}. This kit facilitates the simultaneous operation of two 12-megapixel IMX477 camera modules through a single MIPI CSI-2 camera socket on the Jetson Nano. The kit includes two high-quality camera modules and an Arducam HAT stereo camera, which allows the ArduChip to present a dual camera connection as a single camera recognized by single-board computers. Such a kit is ideal for constructing a stereo camera system for depth vision applications, capitalizing on the benefits of RGBD technology.

\subsection{Audio/Microphone Array}

While the majority of research efforts have centered on the development of video-based solutions, UAV-embedded audio-based localization has garnered comparatively less attention \cite{Basiri2012, Strauss2018}. Nevertheless, the utilization of UAVs equipped with microphone arrays holds the potential to be of paramount importance in locating individuals during emergency situations. This potential becomes particularly evident when video sensors are hindered by factors such as poor lighting conditions (e.g., at night or in fog) or obstacles that restrict their field of view \cite{Deleforge2019}.

When recording audio, unwanted background noise can often degrade the quality of the recording. To address this issue, digital processing is used to enhance the sound, eliminate noise, and retain only the desired audio. This process often involves the creation of multiple recordings from different locations and then comparing these recordings to distinguish between the desired sound source and the various noise sources. Subsequently, the unwanted noise can be removed. A microphone array, consisting of two or more microphones used together, is a valuable tool for identifying and isolating individual sound sources.

The human auditory system serves as a natural example of a microphone array. Human ears allow individuals to determine the approximate direction of a sound source in real time. Without conscious effort, people instinctively know where to turn their heads to hear a voice, where to focus their attention when an unexpected sound occurs, and which way to move to avoid potential threats. This innate ability is possible because humans have two ears, positioned on each side of their head.

Microphones are employed to replicate this sound source recognition process by digitally capturing audio. Sound travels as pressure waves through the air, and just as the human ear has an eardrum that senses changes in nearby air pressure, microphones are equipped with a thin diaphragm that vibrates in response to variations in air pressure. When the diaphragm moves, the coil of wire wound around a magnet also moves, generating an electrical signal that faithfully reproduces the recorded sound. This electronic signal is then stored in a computer for subsequent analysis. In a microphone array, each microphone captures a signal that is stored and can be compared with the signals from other microphones in the array. Each microphone, being at a different distance from the sound source, experiences a unique time delay, a characteristic depicted in the accompanying figure. This time delay information is crucial for sound localization and source separation in microphone array systems.

\begin{figure}[t]
\centering
\includegraphics[width=8.0cm]{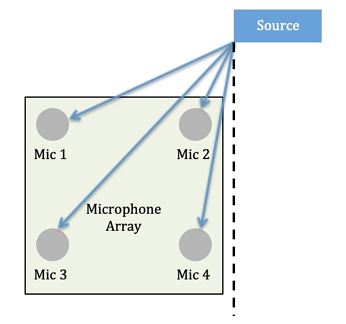}
\caption{Microphone array.}
\label{Audio01}
\end{figure} 

The microphone array system utilizes precise calculations based on time delay differences to determine the location of sound sources. This allows the system to gather valuable information about its environment, including the positions of sound sources and the distinct characteristics of noise-free sounds.

To calculate the location of a sound source, the microphone array employs specialized software. Initially, it estimates the time delay between a primary or main microphone and each of the other microphones in the array. The calculation involves holding the primary audio recording in a fixed position while overlaying a secondary recording on it and repeatedly shifting it across for a brief duration. At each shifting point, the system calculates the difference between the two audio signals. When the difference between two recordings is minimal, it indicates that they are closely aligned in time.

The time delay, defined as the amount by which the secondary microphone recording needs to be shifted to align with the primary recording, is then determined for each secondary microphone. With these time delay measurements, the system can approximate the position of the sound source using geometric principles. By knowing the speed of sound, it can calculate the distance from each microphone to the source based on the time delay. The figure illustrates that a set of distances, starting at the same point and ending at each microphone, only corresponds to one possible source location. Therefore, if the time delay estimates are accurate, the system can pinpoint the source location with high precision.

\begin{figure}[t]
\centering
\includegraphics[width=8.5cm]{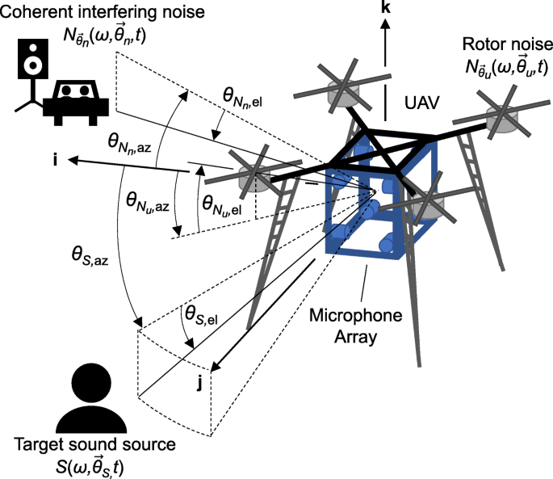}
\caption{Localization based on angle of arrival.}
\label{aoa}
\end{figure}

\begin{figure}[h]
\centering
\includegraphics[width=8.5cm]{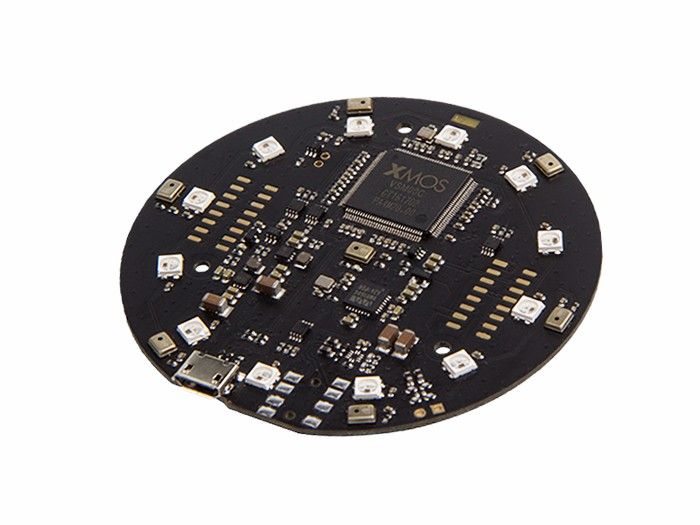}
\vspace{-5mm}
\caption{ReSpeaker.}
\label{ReSpeaker}
\end{figure}

This process is repeated for various combinations of audio recordings, allowing the microphone array to identify multiple sources and their respective locations as shown in Fig. \ref{aoa}.
By doing so, unwanted sounds can be effectively and strategically removed, leaving behind improved sound quality for further analysis and processing. This capability enables the microphone array to enhance sound clarity and precision in various applications. Fig. \ref{ReSpeaker} shows ReSpeaker microphone array that can be suitable for such applications. 

\subsection{Radars}
Intelligent onboard sensors, such as cameras, microphone arrays, and radar, which are particularly effective during the early stages of disaster response. However, radar's unique ability to penetrate through objects and operate in low-visibility conditions makes it indispensable for detecting occluded human subjects, both on and under debris, when other sensing abilities may falter
\cite{Rong2021}. A radar is an electromagnetic sensor utilized for the detection, localization, tracking, and identification of various objects over significant distances. Its operation involves the transmission of electromagnetic energy towards objects, commonly referred to as targets, and the subsequent observation of the echoes reflected from these objects. Radar not only determines the presence, location, and speed of these objects but in certain cases can also ascertain their size and shape. What sets radar apart from optical and infrared sensors is its ability to detect distant objects under adverse weather conditions and accurately measure their distance.

Radar is an ``active'' sensor because it generates its own electromagnetic signals (transmitter) to detect targets. Typically, radar operates in the microwave portion of the electromagnetic spectrum, with frequencies ranging from 400 megahertz (MHz) to 40 gigahertz (GHz), measured in hertz (cycles per second).

UAV radars offer the advantage of working irrespective of weather conditions and lighting, a capability not shared by many electro-optical sensors. They are capable of detecting autonomous drones, while radio frequency sensors rely on intercepting signals transmitted between drones and their human operators. The ease of detecting a UAV with radar depends on its effective scattering area commonly referred to as radar cross section (RCS), which is influenced by the UAV's size and the amount of reflective material it contains. UAVs with larger RCS values can be detected at greater distances.

Ground surveillance radars play a crucial role in scenarios where aircraft operating in civilian airspace must be equipped to detect and avoid other aircraft. This is typically accomplished with human pilots on board manned aircraft. However, for unmanned aerial vehicles, ground-based surveillance and avoidance systems (GBSAA) can be employed. These systems utilize ground surveillance radars to eliminate the need for human correction or the presence of manned chase aircraft. GBSAA surveillance radars can detect movements in airspace and provide real-time tracking based on known GPS coordinates of the radar, as well as distance and angle to the target. This information can be relayed to the UAV operator, enhancing situational awareness for safe flight operations at all times.

\subsection{Infra-Red Thermography}
Infrared thermography involves utilizing a thermal imager to capture radiation (heat) emitted from an object, converting it into temperature data, and presenting an image representing the temperature distribution. These recorded temperature distribution images are known as thermograms and enable the visualization of heat-emitting objects that are not visible to the naked eye. Because all objects emit thermal infrared energy above absolute zero (-459.67 degrees Fahrenheit), thermal imagers can detect and display infrared waves irrespective of ambient lighting conditions. A classic example of this technology is the use of night vision goggles to observe objects in darkness.

An infrared thermometer, in its simplest configuration, consists of a lens that focuses infrared thermal radiation onto a detector, which then translates the radiation energy into a color-coded signal. Infrared thermometers are designed for non-contact temperature measurement, eliminating the need for physical contact with the object being measured. Nowadays, there are various types of infrared thermometers tailored for specific applications. The three most common categories of infrared thermometers include:

    \begin{itemize}
    \item \textbf{Spot infrared thermometers :} A spot infrared thermometer, often resembling a handheld radar device, is utilized for detecting and measuring the temperature at a specific point on the surface of an object. These thermometers are particularly well-suited for applications where it may be challenging to access the target object or when the object operates in extreme conditions. Spot infrared thermometers offer the advantage of providing quick and accurate non-contact temperature readings, making them valuable tools in various industries and scenarios.
    \item \textbf{Infrared scanner systems:}
    Infrared scanning systems, which are designed to scan large areas, find extensive use in manufacturing plants with conveyor systems or continuous processes. These systems excel at scanning and monitoring objects on a conveyor belt or sheets of materials, such as glass or metal, emerging from industrial furnaces. This application is a prime example of how infrared scanning technology can be employed to ensure product quality and process control in manufacturing environments. By capturing temperature data across large surfaces, these systems help maintain product consistency and identify anomalies or defects promptly, making them invaluable tools in industrial settings.
    \item \textbf{Infrared thermal-imaging cameras:}
    Thermal imaging cameras represent an advanced category of radiation thermometers that are employed to measure temperatures at multiple points across a wide area, ultimately producing two-dimensional (2D) thermographic images. These cameras are considerably more complex, both in terms of their hardware and software, compared to spot thermometers. They typically offer live image displays and can be connected to specialized software for in-depth analysis, enhanced accuracy, and comprehensive reporting. Modern thermal imaging cameras are designed to be portable, allowing users to capture temperature data in various settings and conditions. These cameras often feature multiple-color palettes that assist in interpreting temperature differences more effectively.
    \end{itemize}
Infrared thermal imaging cameras provide users with the flexibility to switch between various color palettes, including options like the hot iron palette, black and white palette, and rainbow palette, which aid in distinguishing temperature variations.

\begin{figure}[b]
\centering
\includegraphics[width=8.7cm]{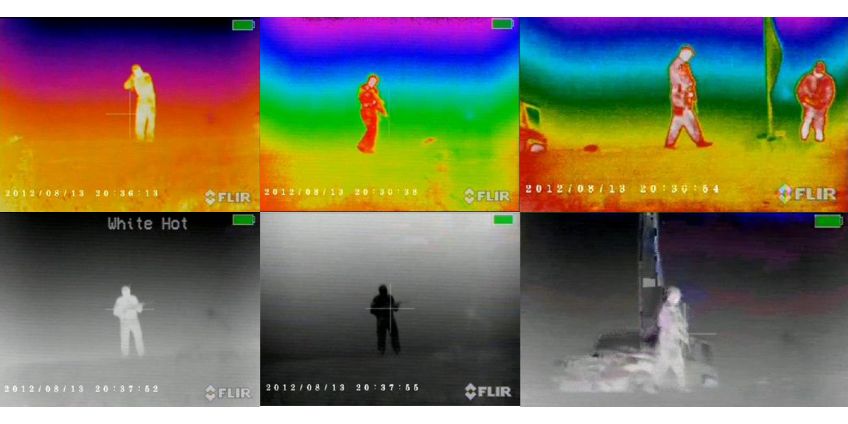}
\caption{Infrared Palette.}
\label{Ifrared}
\end{figure}
\begin{figure}[t]
\centering
\includegraphics[width=6.0cm]{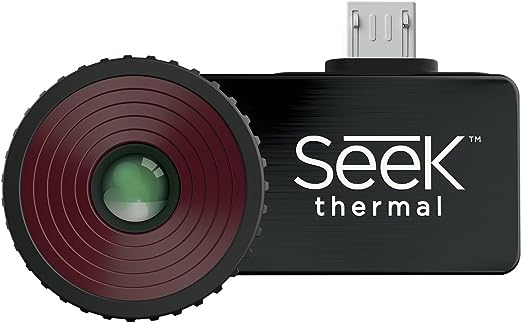}
\caption{Seek Thermal Compact PRO XR.}
\label{SeekImage}
\end{figure}

 
We utilized the Seek Thermal Compact PRO XR camera, shown in Fig. \ref{SeekImage}, which excels at detecting people and animals during low-light conditions, such as dawn and dusk, and can identify them at considerable distances where visible light is insufficient. Additionally, its extended range detection capabilities enhance visual perception and situational awareness.

\section{Advantages and disadvantages of the hardware}
\label{AdvantagesAndDisadvantages}
\subsection{Video Camera}

Arducam IMX477 advantages:

\begin{enumerate}
\item 	The Seek Thermal Compact PRO XR, as mentioned, has the capability to support 12-megapixel digital still images at a resolution of 4056×3040 pixels.
\item	Full resolution at 60fps (normal), 4K2K at 60fps (normal), 1080p at 240fps.
\item	The Arducam IMX477 camera module features a motor that can be controlled through software, enabling more intelligent focusing. This means that users no longer have to manually adjust the camera's focus by physically turning the lens with their hands.
\item	The Arducam IMX477 camera module not only features motorized focusing but also supports autofocus functionality. 
\item	The presence of pins on the camera board dedicated to a mechanically switchable IR filter is a notable feature of the Arducam IMX477 camera module. This switchable IR filter allows for flexibility in applications where both visible light and infrared light are required, either separately or in different conditions.

\end{enumerate}
Arducam IMX477 disadvantages:
\begin{enumerate}
\item	The Arducam IMX477 camera module cannot physically move the camera and lens for remote direction and zoom control. This means that the camera's orientation and zoom level are fixed and cannot be adjusted remotely.
\item	It's important to note that one of the camera modules in the Arducam IMX477 camera kit is designed to work in conjunction with the Camarray HAT (Hardware Attached on Top). This means that this particular camera module cannot be directly connected to the Jetson Nano board alone, and it requires the Camarray HAT for proper functionality.
\item	It's important to note that not all Jetson boards are compatible with the Arducam IMX477 camera module. Currently, the camera module is supported by the Jetson Nano and Xavier NX boards specifically. Other Jetson boards may not have the necessary hardware or software support to work with this particular camera module.

\end{enumerate}

\subsection{Microphone Array}
Advantages:
\begin{itemize}
\item	Identifies each direction of the sound source.
\item	Using the Microphone Array, we can identify the type of sound source.
\item	Undesirable sounds can be intentionally eliminated to a large extent, allowing for a more precise and clear analysis of the improved audio.

\end{itemize}

Limitations:
\begin{itemize}
\item	Microphones introduce challenges when dealing with audio signals, especially in noisy environments.
\item	Extended processing time
\begin{enumerate}
\item	 Real-time audio processing can be demanding in terms of computational resources, especially when dealing with complex signal processing tasks, such as those involved in microphone array applications. 
\item	Performing cross-correlation calculations between multiple microphone signals, especially when using a large number of microphones, can be computationally intensive and time-consuming.	
\end{enumerate}
\end{itemize}

\subsection{Radar}
Benefits:
\begin{itemize}
\item	One of the significant advantages of radar technology is its ability to penetrate various environmental conditions (clouds, fog, haze, and snow) that can hinder or limit the effectiveness of other sensors, such as optical or infrared sensors.  
\item	The radar signal can penetrate insulators (materials that are considered insulating, such as rubber and plastic).
\item	Can determine the target's speed.
\item	Can measure the distance to an object.
\item	Can tell the difference between stationary and moving targets.

\end{itemize}

Limitations:
\begin{itemize}
\item	Short range (200 feet).
\item	The radar can interfere with several objects and media in the air.
\item	Cannot distinguish or resolve multiple targets.
\item	Cannot detect targets covered with conductive material.

\end{itemize}

\subsection{Infra-Red Thermography}
Seek Thermal CompactPRO XR benefits:

\begin{itemize}
\item	Allows you to detect objects in conditions of insufficient visibility.
\item	Allow users to switch between multiple color palettes.
\item	High infrared resolution, detecting objects at a distance of 550 meters.
\item Waterproof, dustproof (thanks to a protective cover).
\item	Manual focus.
\item	Does not require batteries or charging.

\end{itemize}

Seek Thermal CompactPRO XR limitations:
\begin{itemize}
\item	The camera itself is not waterproof or dustproof.
\item	Field of view is only 24°.
\item	Works only with an additional device (for example, a phone).
\item	Limited range.
\end{itemize}

\section{Design challenges for UAV-assisted emergency networks}
\label{DesignChallenges}
\subsection{Synchronization Among UAVs}
UAVs, in general, offer advantages such as rapid deployment, flexible reconfiguration, and improved communication capabilities due to their short-range line-of-sight links. Nevertheless, the deployment of highly mobile and energy-constrained UAVs for wireless communications also brings forth a host of new challenges \cite{Zeng2016}. 

UAVs have become indispensable in modern life, and achieving precise time synchronization is crucial for multi-UAV flight. However, the current approach, which relies on GNSS satellite navigation systems or ground control stations for time synchronization, has limitations and poses significant security risks \cite{Panda2019}. The challenge lies in achieving highly accurate time synchronization for all drones within the network independently, without relying on external time sources.

\begin{figure}[b]
\centering
\includegraphics[width=8.0cm]{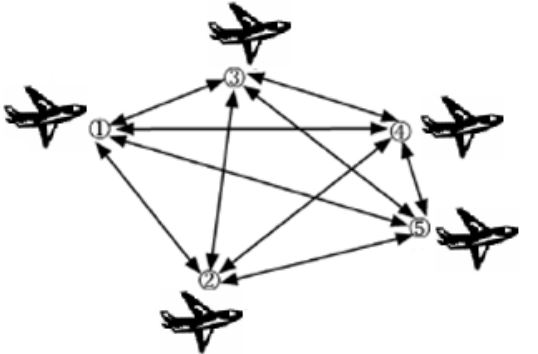}
\caption{UAVs in a coordination.}
\label{uavs}
\end{figure}

In the context of a distributed architecture, such as a large-scale UAV formation, the traditional master-slave time synchronization method is no longer suitable. Instead, the firefly synchronization model, which has ancient origins and has been studied in various fields like biology, chemistry, and mathematics, offers a novel approach to address the issue of distributed time synchronization.
The concept involves each UAV transmitting its current time information \cite{Liu2015}. Once neighboring nodes receive this information, they perform a straightforward arithmetic averaging operation. This calculated average is then used as the countdown for the next transmission. This process is iterated several times until all nodes in the network eventually converge to identical clock values on average. This achieves distributed time synchronization for the entire formation network as depicted in Fig. \ref{uavs}.

In a UAV formation, multiple UAVs are interconnected and can interact with one another \cite{Alsolami2021}. Each UAV serves as a node within the network, establishing a decentralized network structure without a central node. The relative positioning between these nodes is consistent.

Even if one node goes offline or becomes disconnected for various reasons, the remaining nodes can continue to maintain network communication and perform other tasks. This means that the loss of a node within the network does not necessarily disrupt network reconstruction. Additionally, each node in the network can utilize its assigned frequency and propagation code to modulate the propagation spectrum and transmit messages to other neighboring nodes. This transmission process follows a broadcast-style approach, enabling one-to-multipoint communication with multiple access capabilities.

\subsection{UAV Network Security}

In recent years, there has been a notable increase in the malicious use of unmanned aerial vehicles (UAVs). These attacks have become more frequent and can have severe consequences. As a result, various industries and standardization bodies are actively exploring ways to enhance the security of UAV systems and networks \cite{KaiYun2022}.

To address this issue comprehensively, threats and protective measures are categorized based on the characteristics of the first four layers of the OSI model: the physical layer, data link layer, network layer, and transport layer. In order to provide a deeper insight, the security mechanisms under examination are thoroughly assessed in terms of their security requirements and objectives \cite{Erdelj2017}. These objectives encompass aspects such as availability, authentication, authorization, confidentiality, integrity, and non-repudiation.

The classification of security threats according to security requirements is outlined below. We will now proceed to discuss the three most prevalent types of attacks in this context.

\subsection{Active Interfering (Jamming)}
In the context of wireless connections, active interference represents a form of disruption that primarily impacts availability. To combat this type of attack, various strategies are employed, such as frequency-hopping spread spectrum (FHSS) and direct sequence spread spectrum (DSSS).

FHSS operates by swiftly altering the channel frequency within a non-overlapping range while transmitting radio signals. This dynamic channel switching helps minimize interference with adjacent radio channels and reduces the vulnerability to jamming attacks by employing random patterns for channel transitions.

DSSS, on the other hand, combines radio frequency carriers with pseudo-noise digital signals to generate a broad transmission signal. This signal carries a greater amount of information by utilizing a wider bandwidth, enhancing resistance to active interference.

\subsection{Denial-of-Service (DoS)}

Denial-of-Service (DoS) attacks pose a significant threat to the smooth operation of drones. To counteract these attacks, the CoMAD protocol can be employed for mitigation. Initially, redundant data originating from the same source UAV is eliminated. Additionally, any UAVs attempting to access the system with incorrect passwords or unauthorized authentication contexts are restricted from operation.

\subsection{GPS Spoofing}
GPS spoofing presents a serious threat to UAV operations, potentially leading to hijacking. To promptly identify GPS spoofing attacks, the CUSUM algorithm is employed. This algorithm monitors the hit rate and detects GPS spoofing when the time offset surpasses the predefined no-hit zone. For instance, HID-RS deploys rule-based intrusion detection techniques on both UAVs and Ground Control Stations (GCSs) along with corresponding response mechanisms. Typically, attackers emit high-intensity signals (SSI) to gain control of UAVs. Thus, HID-RS equips UAVs with agents to collect SSI data from the source node. The collected SSI data is then assessed against a predetermined SSI threshold to determine the presence of GPS spoofing. The SSI threshold is dynamically adjusted using the Support Vector Machine (SVM) algorithm to maintain effective detection capabilities.

\section{Design Architecture}
\label{DesignArchitecture}
Numerous advanced technology applications, such as employing a hexacopter for various purposes, necessitate a well-structured architecture comprising both hardware and software modules \cite{Hu2021}. In the case of a hexacopter, this layered organization is instrumental in facilitating efficient functionality and task coordination. The design encompasses various hardware and software components that collaboratively operate to accomplish predefined objectives.

\textbf{Hardware Modules}: 
\begin{enumerate}
\item \textbf{Hexacopter Frame and Components:} The hexacopter system centers on a robust yet lightweight carbon fiber frame meticulously engineered to accommodate a range of crucial elements, including motors, propellers, batteries, and the payload.

\item \textbf{Flight Controller:} Serving as the hexacopter's central nervous system, the flight controller assumes responsibility for real-time aircraft stabilization and control. Employing cutting-edge algorithms, it dynamically regulates motor speeds to ensure flight stability. Furthermore, it collects telemetry data on flight status, battery levels, and sensor readings.

\item \textbf{Power Distribution System:} Responsible for efficiently distributing power from the batteries to all components, this system ensures a consistent and reliable electricity supply throughout the hexacopter.

\item \textbf{Motors and Propellers:} Equipped with six high-quality brushless motors, each linked to a propeller, the Hexacopter generates the necessary thrust to achieve flight and stability.

\item \textbf{Battery System:} The Hexacopter relies on a high-capacity lithium-polymer (LiPo) battery to supply the required energy for propulsion and onboard electronics.

\item \textbf{Camera:} The Arducam stereo camera HAT enables the easy integration of these camera modules with single-board computers like the Jetson Nano. It delivers high-resolution imagery, rendering it ideal for applications such as mapping, surveillance, and image-based inspections. Visual data plays a crucial role in tasks like navigation, object recognition, and environmental analysis.

\item \textbf{Thermal Camera:} Thermal sensing represents another crucial aspect of the application layer. It empowers the UAV to identify heat signatures, which prove invaluable for locating living subjects, even in low-light or obscured conditions.

\item \textbf{Voice and Sound Recognition:} An advanced microphone array sensor, combined with voice and sound recognition algorithms, aids in identifying human voices or distress signals. This capability is pivotal in search and rescue operations, facilitating the pinpointing of the locations of survivors or individuals requiring assistance.

\item \textbf{Controller:} The Jetson Nano controller gathers data from a diverse array of sensors integrated into the drone. These sensors encompass vision sensors, thermal sensors, and microphone sensors, each offering unique information about the drone's surroundings. The Nano controller processes this sensor data to make well-informed decisions.
\end{enumerate}

\begin{figure}[h]
			\centering
	\includegraphics[scale=0.8]{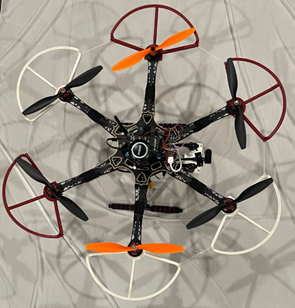}
	\vspace{-0.1cm}
\captionof{figure}{The Helicopter Drone Prototype.}
\label{fig:Helicopter}
\end{figure}

The prototype of the helicopter drone, complete with all sensors and an embedded Jetson Nano minicomputer, is depicted in Fig. \ref{fig:Helicopter}.

\textbf{Software Modules:} 
\begin{enumerate}
\item \textbf{Flight Control Software:} Operating on the PX4 PIX 2.4.8 Flight Controller, the flight control software assumes responsibility for stabilizing the hexacopter, overseeing its movements, and executing commands relayed from the RadioLink AT10 transmitter. This software holds critical significance in guaranteeing stable and precise flight performance.

\item \textbf{Radio Communication Software:} The RadioLink AT10 transmitter establishes communication with the hexacopter through 2.4GHz radio signals. The dedicated software governing this communication system ensures the dependable and real-time transmission of control commands.

\item \textbf{Navigation and Control Algorithms:} Within the hexacopter's architecture, the navigation and control layer encompasses algorithms tasked with functions like path planning, altitude control, and obstacle avoidance. These algorithms collaborate closely with the flight controller to guarantee the safety and efficiency of flight operations.

\item \textbf{User Interface:} Although not explicitly highlighted, user interfaces (UI) or ground control software (GCS) frequently play a pivotal role in hexacopter operations. These interfaces empower users to oversee the hexacopter's status, strategize flight missions, and fine-tune control parameters.

\item \textbf{Sensor Fusion and Data Processing:} Hexacopters come equipped with an array of sensors, encompassing accelerometers, gyroscopes, and potentially supplementary sensors dedicated to functions like obstacle detection and mapping. Sensor fusion algorithms harmonize data derived from these sensors, yielding a holistic perspective of the hexacopter's surroundings.

\item \textbf{Task-Specific Applications:} Tailored to the specific purpose at hand, the hexacopter can feature dedicated software modules. For instance, when employed in aerial photography or mapping tasks, specialized software for camera control and image processing might be incorporated. Conversely, in search and rescue missions, essential software may revolve around object detection and tracking.

\end{enumerate}

\subsection{Loading Capability}

Payload refers to the load that a drone carries, encompassing various items such as cameras, sensors, delivery packages, and other technologies tailored to specific requirements. While a drone's ability to accommodate additional equipment and technologies increases its versatility, it's important to note that carrying heavier payloads results in shorter flight times due to increased power consumption, which depletes the battery more quickly.

\subsection{Flight Time and UAV Charging}
For conducting extensive surveys of a specific area, longer flight durations become imperative. Similarly, in emergency scenarios, the need for frequent facility monitoring arises. Commercial UAVs often face constraints due to limited battery capacity, typically resulting in flight times of less than 1 hour. It's worth noting that the overall drone design significantly influences flight duration. For instance, the Parrot Disco, featuring a fixed-wing single-rotor blade design, offers a continuous flight time of 45 minutes, surpassing the 30-minute continuous flight time of the similarly-weighted Da-Jiang Innovations (DJI) Mavic Pro drone, which employs a multi-rotor quadcopter design. This discrepancy is primarily attributed to the better aerodynamics and reduced thrust requirements of the former design.

\subsection{Handling UAV Failures}
Focusing on the engine control system, flight control systems, and human factors, which collectively account for approximately $80\%$ of UAV failures, offers a promising avenue to enhance reliability compared to other manned systems. It is imperative to keep the cost of UAVs at a minimum while ensuring they can execute their tasks with an acceptable level of safety, reliability, operability, and survivability. There are two primary approaches to enhancing UAV reliability: fault tolerance and failure prevention.

Traditionally, fault tolerance relies on hardware redundancy. These schemes are designed with redundant configurations, often triple or quadruple redundancy, to allow for continued operation in the event of a failure. However, fault tolerance via hardware redundancy introduces challenges such as increased maintenance costs, and additional requirements for space, weight, and power consumption, all of which are critical factors for operational UAVs. In cases where size, weight, and cost constraints are paramount, emphasizing crash prevention becomes a more favorable approach.

While there exists extensive literature on fault tolerance methods applied to UAVs \cite{Erdos2013}, the field of failure prevention techniques is relatively less detailed and somewhat outdated. Various methods, including fault tree analysis (FTA), failure modes and effects analysis (FMEA), and component analysis (CA), are employed in this context. FTA serves as a qualitative assessment method to guide designers, planners, or operators in understanding potential system failures and devising strategies to address their causes. It typically complements FMEA, which involves an exhaustive examination of failure modes for each system component, assessing their impact on system operation. FMEA is most effective when implemented during the design phase, allowing critical failure probabilities to be identified and mitigated early in the process.

The reliability methodologies presented here can be applied to virtually any aspect of UAVs, encompassing flight control systems, power systems, communications, and other critical components, enabling comprehensive reliability analysis.

\section{Future Directions}
\label{FutureDirections}
As the field of unmanned aerial vehicles (UAVs) continues to advance, the potential applications for these remarkable machines are expanding at an unprecedented pace. Looking ahead, we foresee a future where drones will play a pivotal role in addressing environmental disasters and humanitarian crises. Here are some possibilities for the future of UAV technology.

\subsection{Disaster Response and Search-and-Rescue Operations}
One of the most promising directions for UAVs lies in their application in disaster response and search-and-rescue missions. Following earthquakes, hurricanes, or other catastrophic events, drones equipped with advanced sensors and imaging technology can be deployed to locate and assist survivors. Looking to the future, we envision the integration of cutting-edge capabilities, including the ability to analyze sound in disaster-stricken areas.

Consider a scenario in which drones not only capture images but also process audio data from their surroundings. Through the analysis of sound patterns, coupled with image recognition technology, UAVs can identify signs of life, such as cries for help or calls for assistance. This innovative approach has the potential to significantly improve the speed and accuracy of search-and-rescue missions, particularly in noisy environments or situations where visual cues alone may prove insufficient.

\subsection{Real-Time Predictive Analysis}
The future of UAV technology extends beyond immediate disaster response to proactive disaster prevention. By collecting and analyzing real-time data, drones can become invaluable tools for predicting and mitigating environmental threats. For example, drones equipped with sensors can detect the early signs of wildfires, monitor their progression, and even predict their future trajectory.

In practice, this means that a drone patrolling a forest area could identify the presence of smoke or rising temperatures that indicate a potential fire outbreak. Utilizing advanced data analytics and machine learning algorithms, the drone could estimate the likely path of the fire and its rate of spread. This crucial information can then be transmitted in real-time to authorities and communities, enabling faster response and more effective evacuation efforts.

\subsection{Improving Sound Analysis in Noisy Environments}

Enhancing the accuracy of sound analysis in noisy environments remains a critical area for improvement. We anticipate significant advancements in this field as we expand our database and refine our machine-learning models. Through the continuous collection and training of a broader range of sound data, our ultimate aim is to achieve the highest possible accuracy in sound detection, even within the most challenging acoustic environments.

The future of UAV technology holds immense promise, particularly in the domains of disaster response and environmental monitoring. By leveraging the capabilities of sound analysis, real-time predictive systems, and ongoing model refinement, drones have the potential to bring about a revolution in how we address environmental disasters and safeguard communities. As we strive for greater precision, reliability, and adaptability in UAV technology, we remain unwavering in our commitment to pushing the boundaries of innovation and making the world a safer and more resilient place.

\section{Conclusion}
\label{conclusion}
In this article, we present a survey focused on the use of drone-based systems for detecting individuals, with a particular emphasis on identifying human screams and other distress signals. This study holds significant relevance in post-disaster scenarios, where drones traditionally capture aerial images to assess structural damage and determine the extent of the disaster. They also leverage thermal imaging technology to detect body heat signatures, aiding in locating individuals.

We delve into various challenges associated with pinpointing humans through aerial acoustics. Furthermore, we explore the application of signal processing techniques, such as direction of arrival (DOA) based on microphone array signals, to enhance the precision of tracking the source of human noises.

The architectural design of such a rescue system has been thoroughly examined and discussed. Additionally, we draw attention to concerns related to UAVs, including vulnerabilities of the wireless network to attacks, disruptions, and synchronization issues.

Lastly, we touch upon future directions, such as the development of real-time systems and the enhancement of sound analysis in noisy environments. 


\bibliographystyle{IEEEtran}
\bibliography{bibs/main}

\begin{IEEEbiography}[{\includegraphics[width=1in,height=1.45in,clip,keepaspectratio]{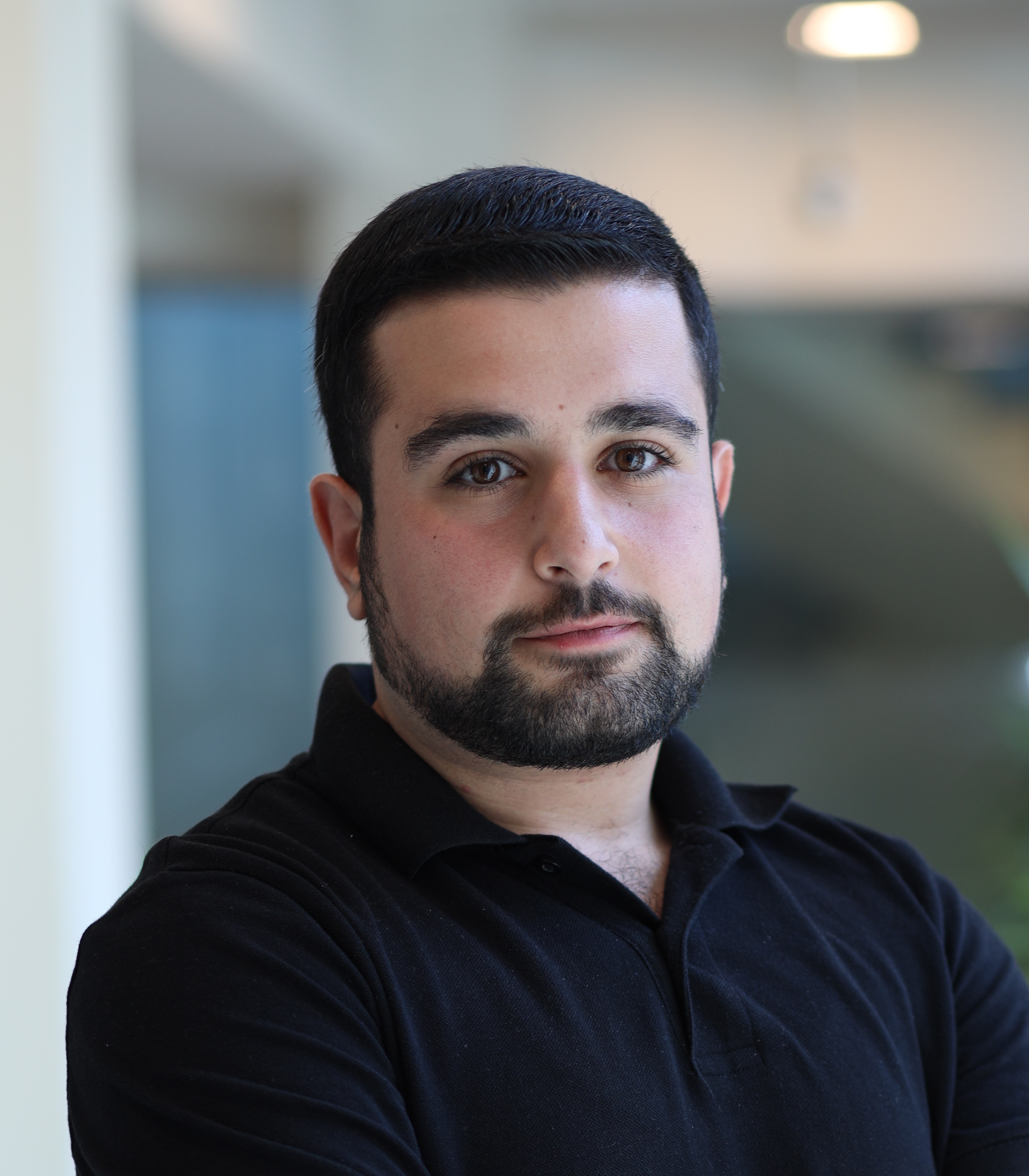}}]{\bf Narek Papyan} received his B.S. degree in Radio-physics from the Yerevan State University in 2019, and his M.S. degree in Wireless Communication and Sensors from Russian-Armenian University in 2023, respectively. Currently, Narek serves as an
Information Security Specialist at the Central Bank of Armenia, where he plays a pivotal role in fortifying the institution’s cybersecurity framework. His responsibilities encompass the implementation and maintenance of rigorous cybersecurity measures to protect sensitive financial data and critical systems.
Narek is a distinguished professional with a strong educational background
and a diverse range of experiences that demonstrate his commitment to information security and technological innovation. Narek's deep learning project 
on ``Change Detection'' resulted in a remarkable $96\%$ accuracy with a minimal $0.055\%$ loss, using a model created by himself on a SiamU-net network with $5$ million parameters. His research interests include machine learning, wireless communications, and cybersecurity.

\end{IEEEbiography}

\begin{IEEEbiography}[{\includegraphics[width=1in,height=1.25in,clip,keepaspectratio]{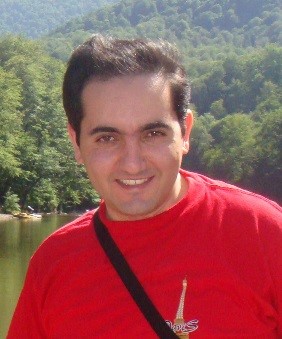}}]{\bf Michel Kulhandjian} (M'18-SM'20) received his B.S. degree in Electronics Engineering and Computer Science (Minor), with ``Summa Cum Laude'' from the American University in Cairo (AUC) in 2005, and the M.S. and Ph.D. degrees in Electrical Engineer from the State University of New York at Buffalo in 2007 and 2012, respectively. He was employed at Alcatel-Lucent, in Ottawa, Ontario, in 2012. In the same year he was appointed as a Research Associate at EION Inc. In 2016 he was appointed as a Research Scientist at the School of Electrical Engineering and Computer Science at the University of Ottawa. He was also employed as a senior embedded software engineer at L3Harris Technologies from 2016 to 2021. Currently, he is a Research Scientist at the Electrical and Computer Engineering Department at Rice University. He received the Natural Science and Engineering Research Council of Canada (NSERC) Industrial R\&D Fellowship (IRDF). 
	
His research interests include wireless multiple access communications, adaptive coded modulation, waveform design for overloaded code-division multiplexing applications, RF and audio fingerprinting, channel coding, space-time coding, adaptive multiuser detection, statistical signal processing, machine learning, covert communications, spread-spectrum steganography, and steganalysis. He actively serves as a member of the Technical Program Committee (TPC) of IEEE WCNC, IEEE GLOBECOM, IEEE ICC, and IEEE VTC, among others. In addition, he serves as an Associate Editor of Annals of Telecommunications Journal, a guest editor for the Journal of Sensor and Actuator Networks (JSAN), and a Review Editor at Frontiers in Communications and Networks. He is a recipient of the best paper award at the 48th Wireless World Research Forum (WWRF) in 2022. 
\end{IEEEbiography}

\begin{IEEEbiography}[{\includegraphics[width=1in,clip,keepaspectratio]{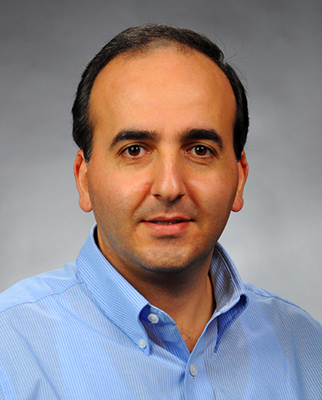}}]{\bf Hovannes Kulhandjian}(S'14-M'15-SM'20) received the
	B.S. degree (magna cum laude) in electronics engineering from The American University in Cairo,
	Cairo, Egypt, in 2008, and the M.S. and Ph.D. degrees in electrical engineering from the State University of New York at Buffalo, Buffalo, NY, USA,
	in 2010 and 2014, respectively.
	From December 2014 to July 2015, he was an
	Associate Research Engineer with the Department of
	Electrical and Computer Engineering, Northeastern
	University, Boston, MA, USA. He is currently an
	Associate Professor with the Department of Electrical and Computer Engineering, California State University, Fresno, Fresno, CA, USA. His current
	research interests include  wireless communications
	and networking, with applications to underwater acoustic communications, visible light communications, and applied machine learning. He has served as a guest editor for IEEE Access - Special Section Journal on Underwater Wireless Communications and Networking. He has also served as a Session Co-Chair for IEEE UComms 2020 and session Chair for ACM WUWNet 2019.
	He actively serves as a member of the Technical Program Committee for ACM and IEEE conferences such as IEEE GLOBECOM 2015-2022, UComms 2022, PIMRC 2020, WD 2019, ACM WUWNet 2019, ICC 2015-2023, among others. 
\end{IEEEbiography}

\begin{IEEEbiography}[{\includegraphics[width=1in,clip,keepaspectratio]{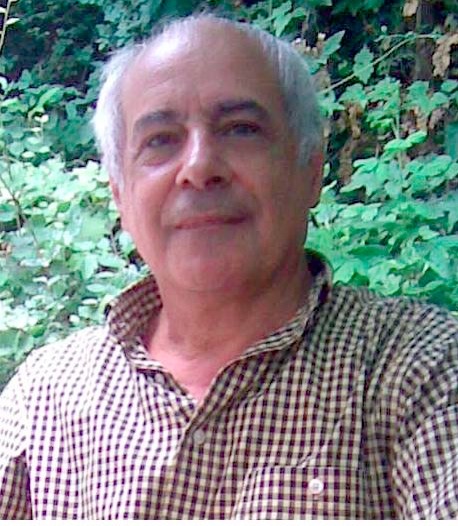}}]{\bf Levon Hakob Aslanyan} received the
	B.S. degree from the Novosibirsk State University in 1968, and the Ph.D. and Doctor of Science (Russian) degrees from the Novosibirsk State University, in 1976 and 1997, respectively. He has been promoted to Professor position since 1997. He has been the corresponding member of the National Academy of Sciences of the Republic of Armenia (NAS RA), since 2014. Currently, he heads the Department of Discrete Modeling, Analysis, and Recognition Technologies at the Institute for Informatics and Automation Problems of the NAS RA. His research interests are in mathematical logic, discrete mathematics, the mathematical theory of pattern recognition, and artificial intelligence. He has made significant contributions in the field of complexity studies of loosely defined Boolean functions, describing sets encompassing all solutions to discrete isoperimetry and discrete tomography problems, advancing logic-combinatorial pattern recognition theory, and contributing to the field of bioinformatics. He authored more than $200$ papers and supervised $2$ doctoral and $14$ Ph.D. students.
\end{IEEEbiography}

\end{document}